\documentclass{elsarticle}

\usepackage{lineno,hyperref}
\usepackage{graphicx,color,amsmath,amssymb}
\journal{Journal of \LaTeX\ Templates}

\begin{document}

\begin{frontmatter}

\title{Inhibition enhances the  coherence in the Jacobi neuronal model
% \tnoteref{mytitlenote}
}

\author[mymainaddress]{Giuseppe D'Onofrio\corref{mycorrespondingauthor}}
%\corref{equa}}
\cortext[mycorrespondingauthor]{Corresponding author}
%\cortext[equa]{eq}
\ead{giuseppe.donofrio@unito.it}

\author[mysecondaryaddress]{Petr Lansky}
\ead{lansky@biomed.cas.cz}

\author[mythirdaddress]{Massimiliano Tamborrino}
\ead{massimiliano.tamborrino@jku.at}

\address[mymainaddress]{Dipartimento di Matematica \lq\lq G. Peano", Universit\`a degli Studi di Torino, Via Carlo Alberto 10, 10123 Torino, Italy}
\address[mysecondaryaddress]{Institute of Physiology of the Czech Academy of Sciences, 
	Videnska 1083, 14220 Prague 4, Czech Republic }
\address[mythirdaddress]{Johannes Kepler University Linz, 
Altenbergerstra\ss e 69, 4040 Linz, Austria  }

\begin{abstract}

The output signal is examined for the Jacobi neuronal model which is characterized by input-dependent multiplicative noise. The dependence of the noise on the rate of inhibition turns out to be of primary importance to observe maxima both in the output firing rate and in the diffusion coefficient of the spike count and, simultaneously, a minimum in the coefficient of  variation (Fano factor). Moreover, we observe that an increment of the rate of inhibition can increase the degree of coherence computed from the power spectrum. This means that inhibition can enhance the coherence and thus the information transmission between the input and the output in this neuronal model. Finally, we stress that 
%the aforementioned measures of spike variability 
the firing rate, the coefficient of variation and the diffusion coefficient of the spike count cannot be used as the only indicator of coherence resonance without considering the power spectrum.

\end{abstract}


\end{frontmatter}

%\linenumbers

%%%%%%%%%%%%%%%%%%%%%%%%%%%%%%%%%%%%%%%%
%%%%%%%%%%%%%%%%%%%%%%%%%%%%%%%%%%%%%%%%%%%
\section{\label{section_1}
Introduction}
%%%%%%%%%%%%%%%%%%%%%%%%%%%%%%%%%%%%%%%%%%%
%%%%%%%%%%%%%%%%%%%%%%%%%%%%%%%%%%%%%%%%%%%

Nonlinear dynamical systems are often strongly influenced by different sources of noise. The response of these systems to random fluctuations has attracted large attention because, differently from the typical assumption that noise can hinder or deteriorate the signal transmission, it has been observed that noise can sometimes improve the information processing both in theoretical models and in experiments \cite{benefits_noise,lind2004,cecchi,levakova,naud}. 
Mathematical models in neuroscience are one of the most prominent examples for which the noise is of primary importance or even a part of the signal itself rather than a source of inefficiency and unpredictability. In this article, we contribute to the discussion on this topic by studying the effects of a multiplicative noise on the performance of a single neuronal model using the analytical approach. 

Typical examples of investigating the effect of noise are numerous studies on the so-called stochastic resonance
 \citep{abbott,SR81,jung_hanggi,gammaitoni,herrmann2013}.
Broadly speaking, the stochastic resonance is observed when increasing the level of the noise improves the signal transmission or detection performance, instead of deteriorating it. The term is traditionally reserved for studying the periodic signals.
% So, the question is if the presence of a weak periodic input within a nonlinear dynamical system can be inferred from the response of the system and if an optimal level of noise exists.
However, it is also natural to ask whether the noise optimizes the information transmission via small aperiodic signals. Then, contrary to the usual setup of stochastic resonance, no external periodic driving is assumed and the coherence that appears as a nonlinear response of the system to the input signal or to a purely noisy excitation  is called aperiodic resonance \cite{collins} or {\it coherence resonance} \citep{gang93,pikovsky,neiman}.
What happens is that for both small and large noise amplitudes, the noise-excited activity appears to be rather irregular, while for moderate noise relatively coherent outputs are observed. Then, the information on the input can be inferred from the available observation of the output. In case of an aperiodic input the word resonance can be misleading, so McDonnell and Ward \cite{benefits_noise} suggested  using the term { \it stochastic facilitation}.

Several specific measures are employed to quantify the above mentioned effects of the noise. For example, the characteristic correlation time is used in \cite{pikovsky}, for the FitzHugh-Nagumo model. Other examples are the cross-correlation coefficient,\cite{collins}, for the integrate-and-fire or Hodgkin-Huxley neuron models or the mutual information between the input and the output, \cite{deco}. 
However, the most commonly used measures are evaluated from the power spectrum both for single neurons \cite{gang93,neiman,mcdonn,pakdaman,lindner2002,guantes,BAUERMANN2019}  and recently for neuronal networks \cite{andreev,wang2018,xie2018}.
The metric we use here as an indicator of the stochastic facilitation is the { \it degree of coherence}, $\beta$.  It is  based on the power spectrum and it is directly related to the Fisher information. In particular, using the Shannon's formula, the total amount of information, $I$, contained in the neuronal output is proportional to $\beta$ (for details see \cite{stemmler})
$$I \propto \int \log_2 [1+\beta]d\omega.$$

The previous  examples of stochastic resonance, coherence resonance or stochastic facilitation are mostly based on the manipulation of the noise component of the system. However, neurons and consequently their models are characterized by the rather intuitive property that the noise amplitude is signal dependent \cite{cecchi,faisal,lanskysac,sac2002,Greenwood2005}.
The signal is composed of excitation and inhibition and it is not so intuitive that despite the integrated level of the signal (sum of excitation and inhibition) is kept constant, due to the manipulation with its components, the noise can be attenuated or enhanced. This may lead to seemingly paradoxical results and the mechanism is also used here. We analyze the dependence of the output on the rate of inhibitory inputs, showing that a small contribution of inhibition can enhance the degree of coherence and thus improving the coding performance. The employed model is based on the Jacobi diffusion \cite{karlin1981second} and
it represents a good compromise between mathematical tractability and biological accuracy.
For the considered model, it is shown in \cite{DTL_jacobi}, that the dependence of the parameters on the rate of inhibition is of primary importance to observe a change in the slope of the response curves. This dependence also affects the variability of the output as reflected by the coefficient of variation, which often takes values larger than one and is not always a monotonic function of the rate of excitation.

The paper is organized as follows. We summarize the Jacobi neuronal model and we recall the relevant mathematical tools in Section  \ref{section_2}. The measures of coherence are listed together with their interpretation and relation with the statistical moments of the first-passage time in Section  \ref{section_3}. These methods of coherence quantification are used to obtain the main results in Section \ref{section_4}  and are then discussed in Section \ref{section_5}.

%%%%%%%%%%%%%%%%%%%%%%%%%%%%%%%%%%%%%%%%%%%
%%%%%%%%%%%%%%%%%%%%%%%%%%%%%%%%%%%%%%%%%%%
\section{The Jacobi neuronal model}
%%%%%%%%%%%%%%%%%%%%%%%%%%%%%%%%%%%%%%%%%%%%
%%%%%%%%%%%%%%%%%%%%%%%%%%%%%%%%%%%%%%%%%%%%
\label{section_2}
The Jacobi neuronal model, $X_t$,  describes the evolution of the neuronal membrane depolarization between two consecutive spikes and is defined by the following stochastic differential equation \cite{DTL_jacobi}, \cite{lanska94}, \cite{longtin_bulsara}
\begin{equation}
\label{Lanska_before}
dX_t=\left(-\frac{X_t}{\tau} +\mu(V_E-X_t)+\nu(X_t-V_I) \right)dt+
\sigma\sqrt{(V_E-X_t)(X_t-V_I)}dW_t, 
\end{equation}
where $ X_0=0$ and 
$\tau > 0$ is the membrane time constant taking into account
the spontaneous voltage decay toward the resting potential
(set equal to zero here) 
in the absence of inputs, $\mu$ and $\nu$. 
The two constants $V_I < 0 < V_E$
 are the inhibitory and excitatory reversal potentials,  respectively.
Here $W_t$ is a standard Wiener process and the diffusion coefficient $\sigma>0$ controls the amplitude of the noise.  
Eq.\eqref{Lanska_before} is obtained in  \cite{lanska87} as a diffusion approximation of a Stein's model with reversal potentials.
In that model two independent homogeneous
Poisson processes represent
the excitatory and inhibitory neuronal inputs, with intensities  $\lambda_E$ and $\lambda_I$, respectively. They describe the arrival of excitatory and inhibitory postsynaptic potentials and are such that the input parameters are
\begin{equation}
\label{input_par}
\mu=e \lambda_E, \qquad \nu=i \lambda_I, 
\end{equation}
where
$i$ and $e$ are constants such that $ - 1 < i < 0 < e < 1$.
The noise amplitude, $\sigma$, is assumed to depend linearly on the input rates
through the constant  $\epsilon > 0$ in the following way
\begin{equation}
\label{input_par2}
\sigma^2=(\lambda_E+\lambda_I)\epsilon.
\end{equation}
Relations \eqref{input_par} and \eqref{input_par2} connect the mathematically tractable but abstract description  \eqref{Lanska_before}, to more biophysical based models as the Stein's one \cite{stein1965,tuckwell_vol2,musila} or the conductance-based models \cite{ric2004, koko, cessac}.
It is obvious from Eqs. \eqref{input_par} and \eqref{input_par2} that with increased input the noise amplitude also increases. In addition, combined with Eq.\eqref{Lanska_before}, it implies that the effect of the input is state-dependent, i.e, the changes in the depolarization 
decrease if $X_t$ approaches $V_I$ or $V_E$, and that the
process is confined in the interval $(V_I, V_E)$.
Throughout the paper, the underlying parameters are chosen
to meet the following condition
\begin{equation}
\label{boundary_cond}
\sigma^2
%=\epsilon(\lambda_E+\lambda_I)
<-\frac{2V_I}{\tau(V_E-V_I)},
\end{equation}
that guarantees that $V_I$ and $V_E$ are entrance boundaries, i.e
the process $X_t$ cannot
reach them in finite time.

To simplify the notation for further calculations, it is convenient to rescale the process $X_t$ in the interval $(0,1)$. Using the transformation $y=\frac{x-V_I}{V_E-V_I}$ in Eq.\eqref{Lanska_before} we obtain
\begin{equation}
\label{Lanska}
dY_t=(-a Y_t+b)dt+\sigma\sqrt{Y_t(1-Y_t)}dW_t, \quad Y_0=y_0,
\end{equation}
with
\begin{equation}
\label{alphabeta}
a=\frac{1}{\tau}+\mu-\nu, \qquad b=\mu-\frac{V_I}{\tau(V_E-V_I)},
\qquad y_0=-\frac{V_I}{V_E-V_I}.
\end{equation}

In accordance
with the model, the spikes of the neuron under study are generated when the process $Y_t$ crosses a voltage threshold $S=\frac{S_0-V_I}{V_E-V_I}$, with $V_I <S_0< V_E$  and $0<y_0<S<1$ for the first time,  the so-called first-passage time (FPT). The process is reset to the starting point $y_0$ after the spike and the evolution starts anew. This reset condition introduces a nonlinearity in the dynamics and guarantees that the interspike intervals (ISIs) form a renewal process. 
In this case, the
ISIs are independent and identically distributed as the FPT, denoted here by $T$ and
defined as
$$T:= \inf \{t\geq 0: Y_t\geq S| 0<y_0<S<1 \}= \inf \{t\geq 0: X_t\geq S_0| V_I<x_0<S_0<V_E \},$$
with probability density function  $g(t)$.
The Laplace transform of  $T$, i.e.,  $g^{*}(\xi):=\mathbb{E}[e^{-\xi T}]=\int_{0}^{\infty}e^{-\xi t}g(t)dt, \xi>0$, is used to calculate the moments of $T$ and the power spectral density and is given by (\cite{lanska94})
\begin{equation}
\label{Lapl_bo3}
g^*(\xi)=\frac{{}_2F_1\left(k(\xi),\theta(\xi);\gamma;y_0\right)}
{{}_2F_1\left(k(\xi),\theta(\xi),\gamma;S\right)},
\end{equation}
where
\begin{equation}
k(\xi)=\frac{2 \xi}{\theta \sigma^2}, \qquad \theta(\xi)=\frac{2a -\sigma^2-\sqrt{(\sigma^2-2 a)^2-8\xi \sigma^2}}{2 \sigma^2}, \qquad \gamma=\frac{2b}{\sigma^2},
 \nonumber
\end{equation} 
and ${}_2F_1$ denotes the Gaussian hypergeometric function.

%%%%%%%%%%%%%%%%%%%%%%%%%%%%%%%%%%%%%%%%%%%%
%%%%%%%%%%%%%%%%%%%%%%%%%%%%%%%%%%%%%%%%%%%%
\section{Methods for coherence quantification}
%%%%%%%%%%%%%%%%%%%%%%%%%%%%%%%%%%%%%%%%%%%%
%%%%%%%%%%%%%%%%%%%%%%%%%%%%%%%%%%%%%%%%%%%%
%
\label{section_3}
The common quantities used in the studies on the neuronal firing activity are the firing rate and the variability of the ISIs.
The firing rate, $f$, is usually defined as the inverse of the mean FPT, i.e $f=1/\mathbb{E}(T)$  \cite{kostalstiber}.
The variability is often characterized by the coefficient of variation of $T$, $\textrm{CV}$, that is the
ratio between the standard deviation and the mean of $T$, i.e.,
$\textrm{CV}=\sqrt{\textrm{Var}(T)}/\mathbb{E}(T)$.
The quantity $ (\textrm{CV})^2$ is equal to the Fano factor \cite{cox_renewal,rajdl_fano,gusella} (also called index of dispersion) for the renewal process. Its value is often used as an exclusive sign of coherence resonance \cite{pikovsky,horikawa,zhong}.
In addition to these quantities, we recall the definition of 
the (effective) diffusion coefficient of the spike count, $D_{eff}$, \citep{cox_renewal}.
% It determines how fast the variance of the spike count grows with respect to . 
For a renewal process, it can be expressed by the moments of  $T$, namely \cite{lind2004,guantes} 
\begin{equation}\label{deff}
D_{eff}=\frac{1}{2}\frac{\textrm{Var}(T)}{\mathbb E(T)^3}=\frac{1}{2} (\textrm{CV})^2  f.
\end{equation}
It determines how fast the variance of $T$ grows with respect to the cube of the mean of $T$.

\begin{figure}[th!]
\begin{center}
 \includegraphics[width=.68\textwidth]{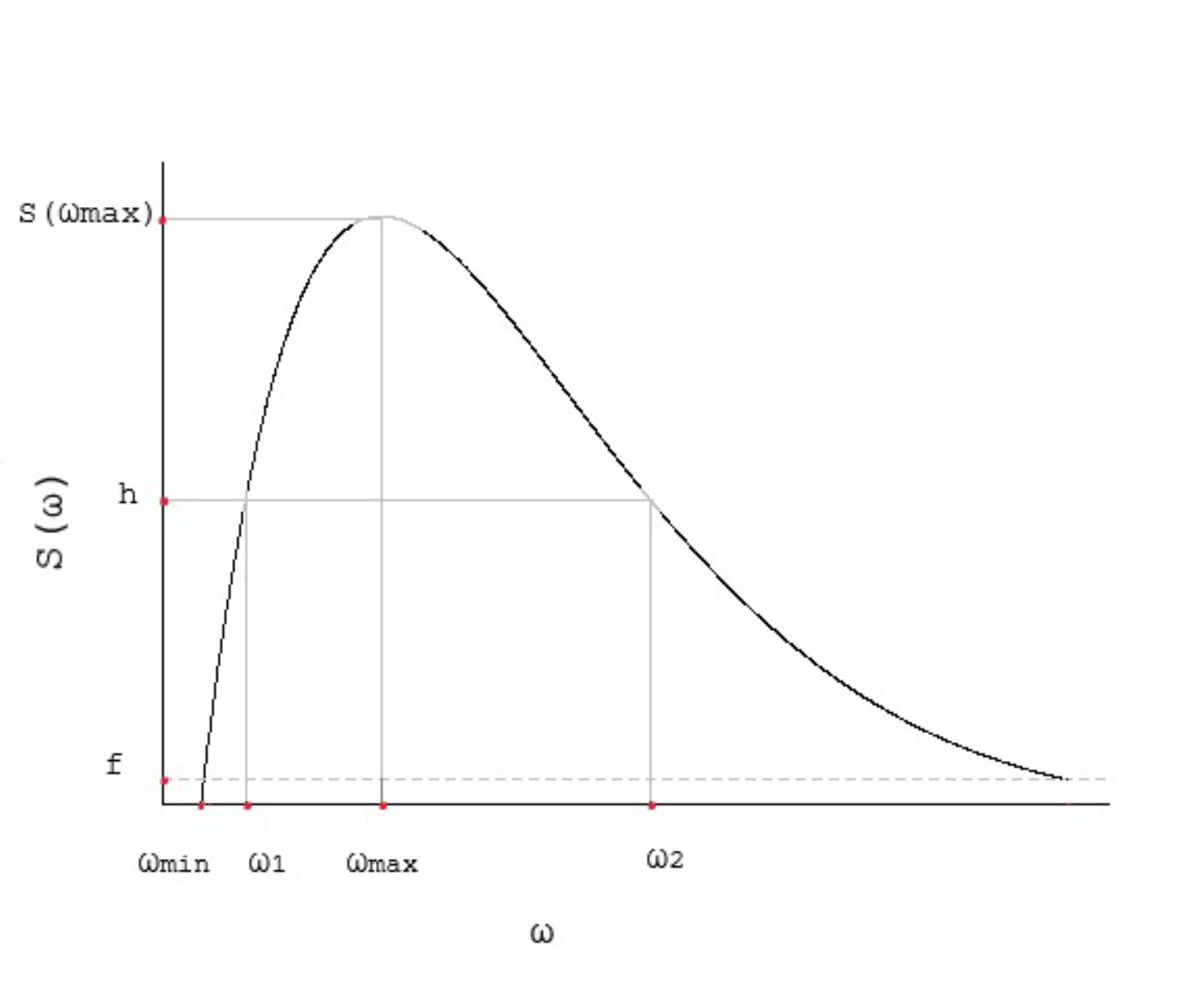}
 \caption{Illustration of the quantities defined in Eq.\eqref{beta_def}. The dashed grey line indicates the firing rate $f$ that is the high-frequency limit of the power spectrum (cf. Eq.\eqref{low_hig_freq}). The value $h$ is the half distance of the maximum of the spectrum, $S(\omega_{max})$, to $f$.  The maximum  $S(\omega_{max})$ must be found after an eventual relative minimum $S(\omega_{min})$.
Note that the power spectrum  for $\omega$ close to zero is either below or above $f$ depending on whether the $\textrm{CV}$  is smaller or greater than $1$, respectively (cf. Eq.\eqref{low_hig_freq}).}
 \label{Fig_beta_def}
 \end{center}
\end{figure}

One of the methods to detect the presence of coherence is the analysis of  the power spectral density  given by
\begin{equation}
\label{power_sp}
S(\omega)=\frac{1-|\rho(\omega)|^2}{|1-\rho(\omega)|^2}f,
\end{equation}
where $\rho(\omega)$ is the Fourier transform of
the  ISI density function, $g(t)$.
The spectrum at vanishing and infinite frequency is
related to the above quantities as follows, \cite{lindner2002}
\begin{equation}\label{low_hig_freq}
\lim_{\omega \rightarrow 0}S(\omega)=2 D_{eff}=(\textrm{CV})^2f, \qquad 
\lim_{\omega \rightarrow \infty}S(\omega)=f.
\end{equation}
 The neuron may possess a noise-induced
eigenfrequency, which appears as a peak in the spectrum, meaning that there is  a preferred frequency.
The size of the peak can be quantified by the degree of coherence, $\beta$, 
which is the ratio of peak height to the relative width
with respect to the position of the maximum of the power spectrum \cite{pakdaman}.
% at which the spectrum decays to a fraction  of its
%maximal value 
%%
%\begin{equation}
%\beta=\frac{S(f_{max})}{\Delta f}f_{max}
%\end{equation}
%%
Here we choose the width at one half of the maximum as in \cite{lindner2002}, that means the half-width of the respective peak over $f$ (see Fig.\ref{Fig_beta_def})
and we pose
$$h:=\frac{S(\omega_{max})+f}{2},$$
where $S(\omega_{max})$ is the maximum value of the power spectrum.
So we use the following version of the degree of coherence 
\begin{equation}\label{beta_def}
\beta=\frac{S(\omega_{max})-f}{\omega_2-\omega_1}\omega_{max},
\end{equation}
 where
%$\Delta \omega=\omega_2-\omega_1$ with
%
\begin{eqnarray*}
\omega_1=\min_{\omega_{min}\leq \omega \leq \omega_{max}} \left\{S(\omega)\geq h \right\},  \qquad
\omega_2=\max_{\omega\geq \omega_{max}} \left\{S(\omega)\geq h \right\},
\end{eqnarray*}
and $S(\omega_{min})$ is the eventual relative minimum value  of the power spectrum attained just before the absolute maximum value, $S(\omega_{max})$. 
%A maximum of the degree of coherence, $\beta$, with respect to noise intensity is considered as a signature of coherence resonance.
The degree of coherence \eqref{beta_def}  often goes under the name of  signal-to-noise ratio (SNR)\citep{gang93}, or coherent SNR \cite{neiman},
that measures how well a spectral peak is expressed with respect to the value of the background noise \cite{lind2004}.

%

%%%%%%%%%%%%%%%%%%%%%%%%
\section{Results}
\begin{figure}[th!]
 \includegraphics[width=.33\textwidth]{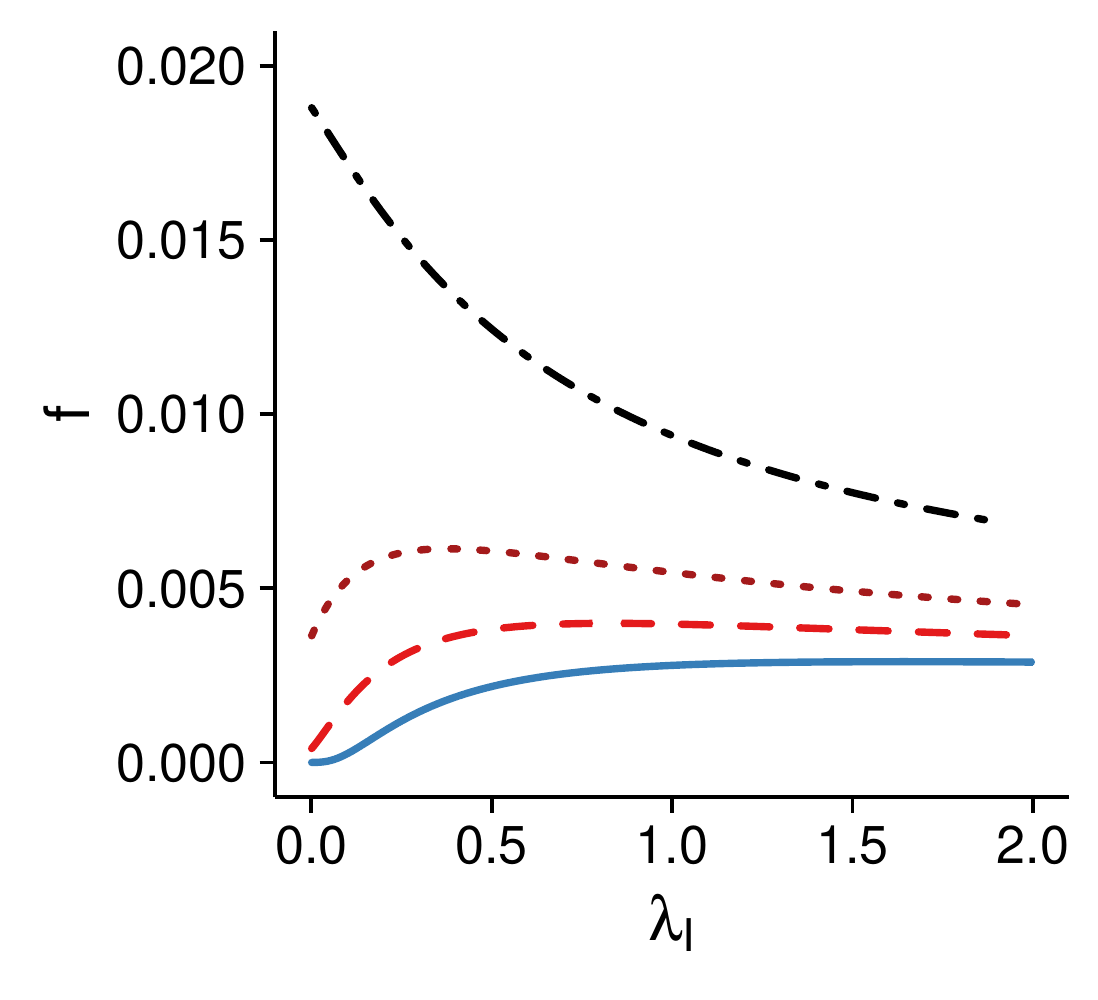}
  \includegraphics[width=.33\textwidth]{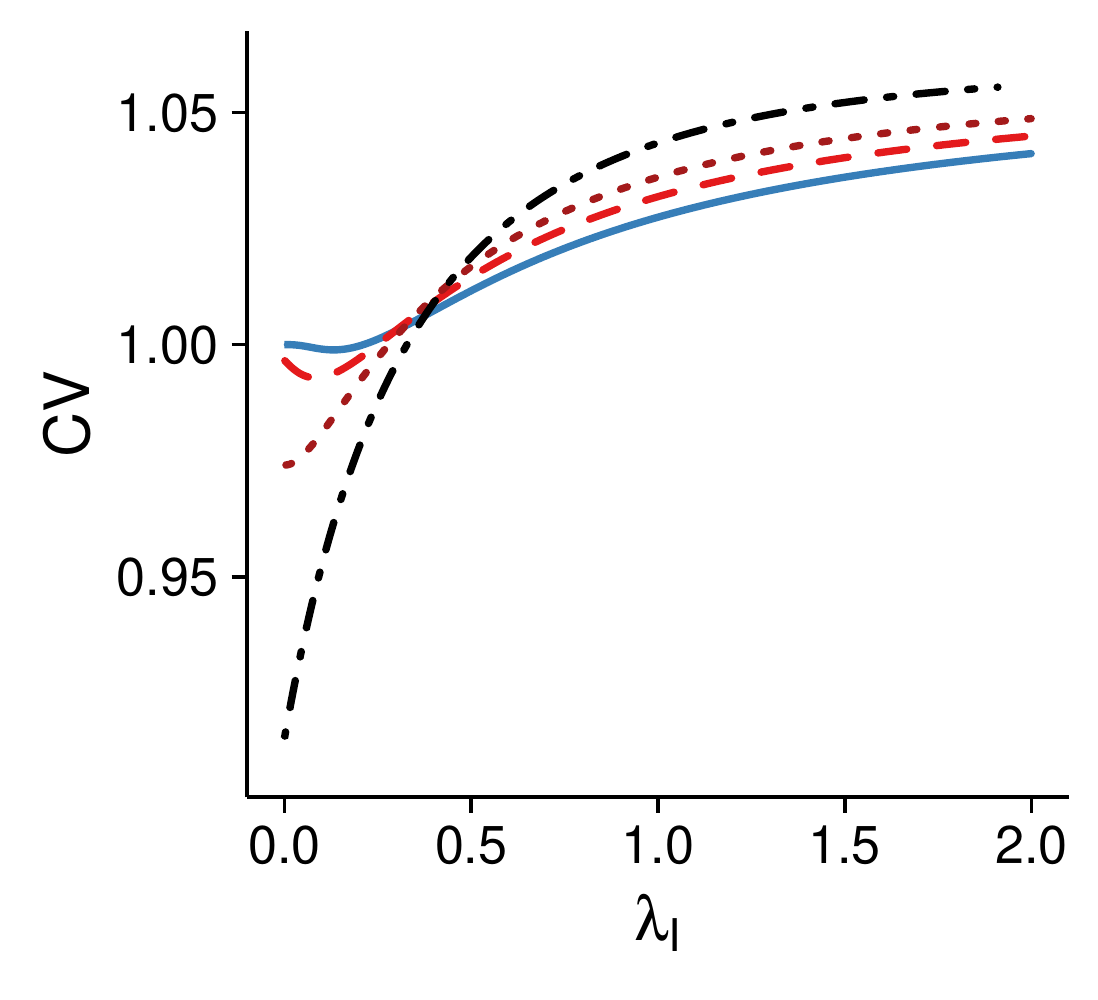}
  \includegraphics[width=.32\textwidth]{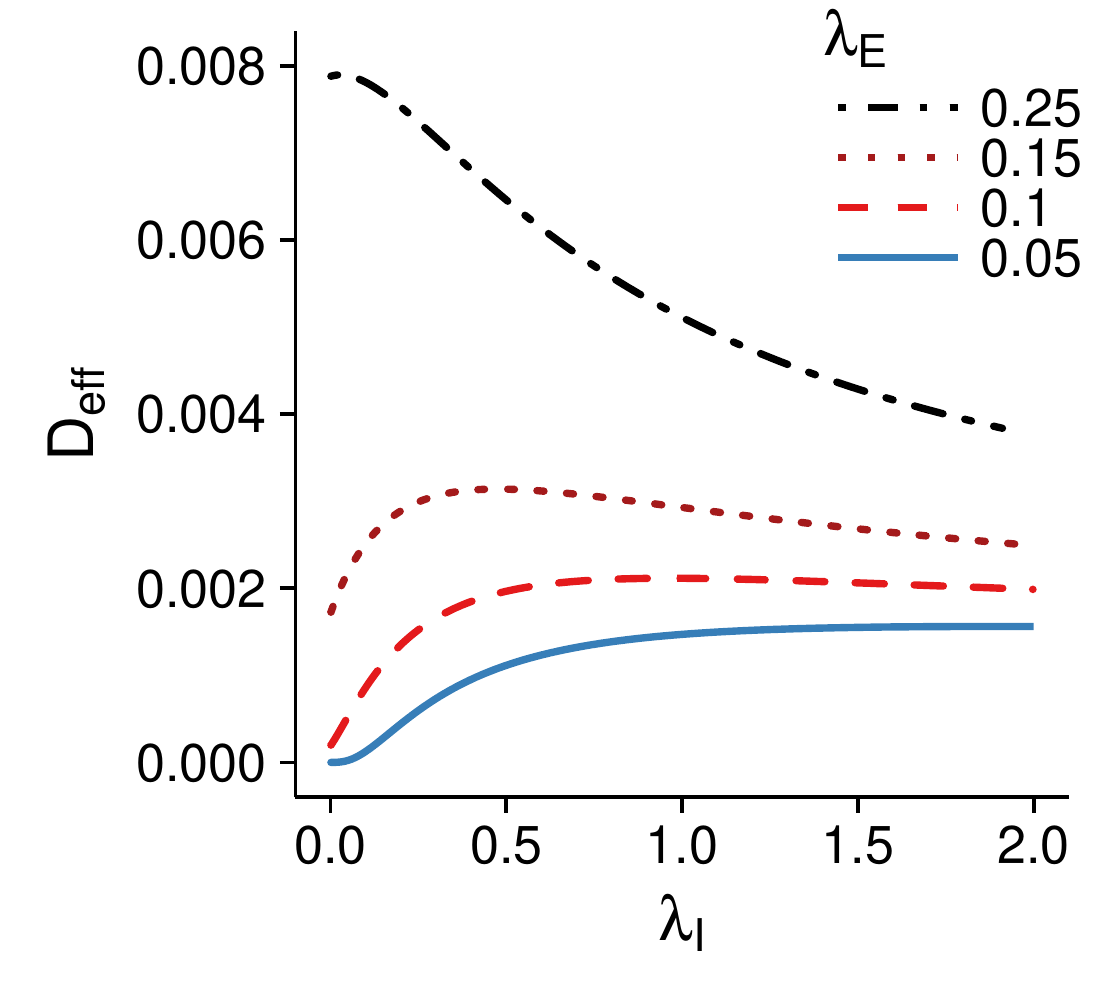}
 \caption{The firing rate, the coefficient of variation of the interspike intervals and the diffusion coefficient of the spike count for the Jacobi neuronal model \eqref{Lanska_before} as a function of the inhibitory rate, $\lambda_I$, for different values of the excitatory inputs rate, $\lambda_E$ (in the legends).
 The values of the other parameters are chosen as in  Ref.\cite{lanska94} :  $S_0=10$ mV, $x_0=0$ mV,  $V_I=-10$ mV, $V_E=100$ mV, $e=0.02$, $i=-0.2$, $\tau=5.8$ ms and $\epsilon=0.0145$.}
 \label{FigET_nu}
\end{figure}

%%%%%%%%%%%%%%%%%%%%%%%
\label{section_4}
We analyze the above quantitative measures of the neuronal firing for the Jacobi model \eqref{Lanska_before} in dependence
 on the rate of the  inhibitory input, $\lambda_I$.
Due to the interrelationship among the parameters in Eqs.\eqref{Lanska_before}-\eqref{input_par2}, some counterintuitive effects are observed. 
The firing rate, $f$, the coefficient of variation, $\textrm{CV}$, and the diffusion coefficient $D_{eff}$  are shown in Fig.\ref{FigET_nu} as a function of $\lambda_I$, for different values of the excitatory rate, $\lambda_E$. 
The firing rate, $f$, is not always decreasing for increasing inhibition. For small values of $\lambda_E$ it can increase with $\lambda_I$ as a result of the form of the noise (cf. Eq.\eqref{input_par}; a detailed discussion is given in Ref.\cite{DTL_jacobi}).
 We note that the CVs show minima and attain values above $1$ for $\lambda_I$ larger than $0.5$ ms$^{-1}$.
The behavior of $D_{eff}$ is similar to that of $f$ because, for the selected parameters,  $\textrm{CV}\approx 1$ (cf. Eq.\eqref{deff}).
For  values of $\lambda_I$ and $\lambda_E$ close to zero, the $D_{eff}$ is almost zero because the neuron is practically silent. 
The maximal values of the $D_{eff}$ are achieved at small values of excitatory rate, $\lambda_E$,  but with   $\lambda_I$ strong enough to produce spikes.
For stronger inhibitory inputs the diffusion coefficient decreases, meaning that the $\textrm{CV}$  is relatively stable with respect to the firing rate.

\begin{figure}[th!]
 \includegraphics[width=.48\textwidth]{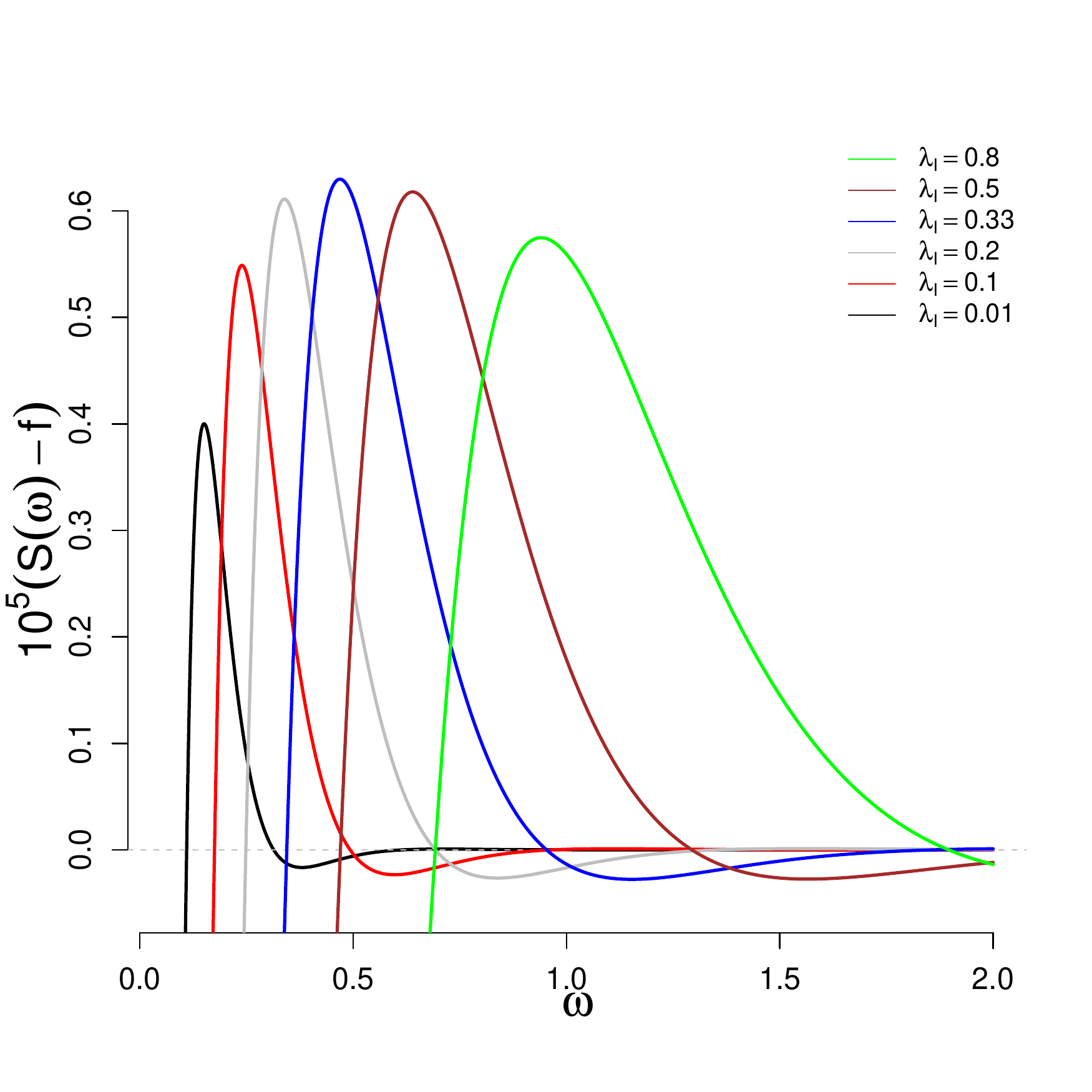}
 \includegraphics[width=.48\textwidth]{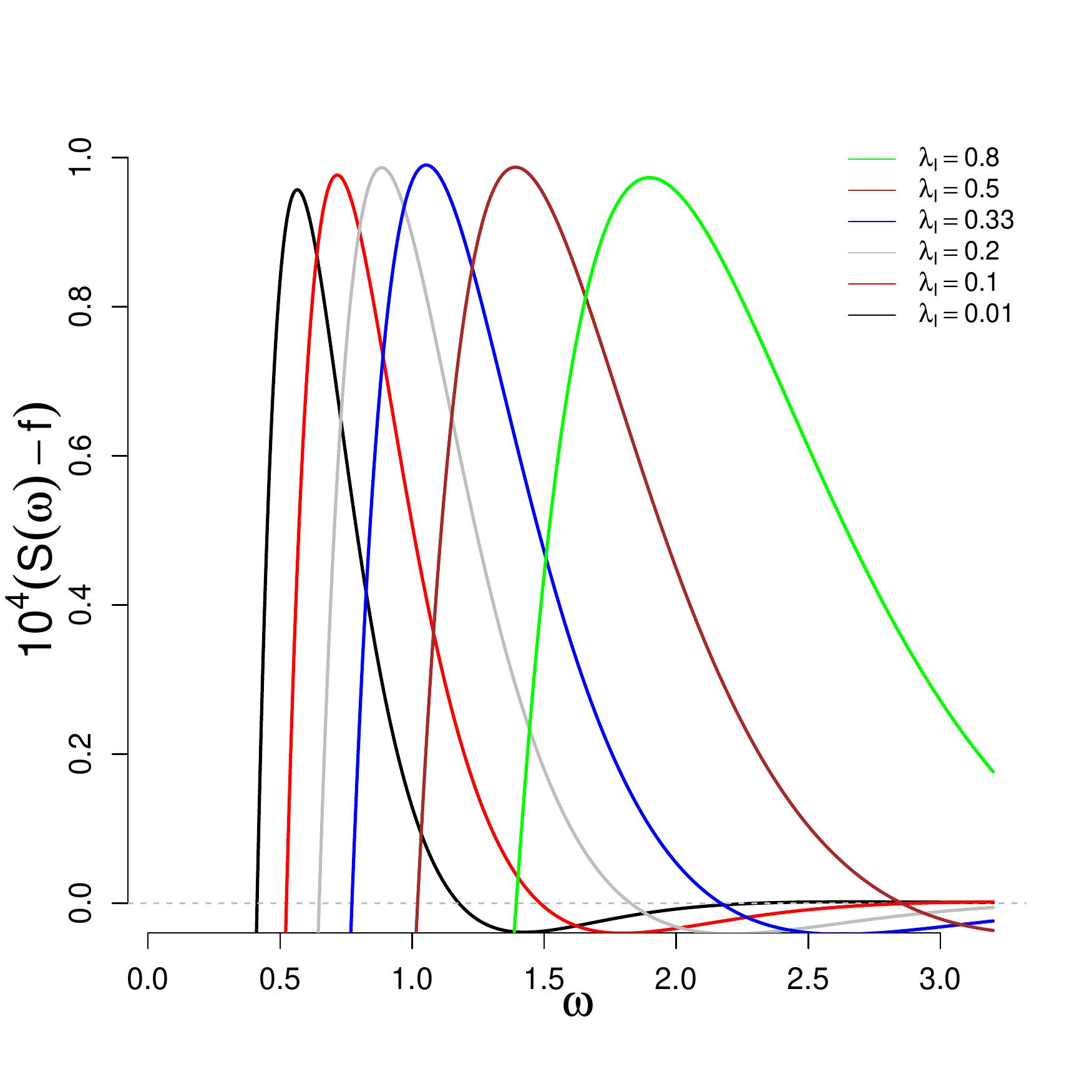} \caption{Rescaled power spectra of the Jacobi neuronal model from Eq.\eqref{power_sp} for increasing values of $\lambda_I$ (from left to right), $\epsilon=0.0145$, $\lambda_E=0.15$ ms$^{-1}$, $\tau=5.8$ ms (left); $\epsilon=0.025$, $\lambda_E=0.34$ ms$^{-1}$, $\tau=3$ ms (right). The other parametres are the same as in Fig.\ref{FigET_nu}. Peaks in correspondence of certain frequencies are clearly visible.
 For increasing values of $\lambda_I$,  the peaks become more prominent. For $\lambda_I > 0.33$ ms$^{-1}$ the peaks decrease their amplitude.  }
 \label{FigPowSpe}
\end{figure}

Evaluating the Laplace  transform in Eq.\eqref{Lapl_bo3} with imaginary argument $\xi = 2 \pi i\omega$, from Eq.\eqref{power_sp} we get the following formula for the power spectrum
\begin{equation}
\label{power_spectrum}
S(\omega)=\frac{1}{\mathbb E(T)}\frac{|{}_2F_1(k(2 \pi i f),\theta(2 \pi i f);\gamma;S)|^2-|{}_2F_1(k(2 \pi i f),\theta(2 \pi i f);\gamma;y_0)|^2}{|{}_2F_1(k(2 \pi i f),\theta(2 \pi i f);\gamma;S)-{}_2F_1(k(2 \pi i f),\theta(2 \pi i f);\gamma;y_0)|^2}.
\end{equation}
Eq.\eqref{power_spectrum} is implemented numerically in the computing environment R \cite{R}.
We show the shifted power spectra, $S(\omega)-f$, to emphasize the height of the peaks in Fig.\ref{FigPowSpe}.
We again choose $\epsilon=0.0145$, $\tau=5.8$ ms,  as in \cite{lanska94} and additionally  $\epsilon=0.025$, $\tau=3$ ms, values also in the physiological range and  satisfying Eq.\eqref{boundary_cond}.
The power spectrum  shows  sharp peaks as a sign of coherence.
In both cases, for increasing values of $\lambda_I$ (from left to right in the figures), the peaks become more prominent up to a certain value of the inhibitory rate and then they start to reduce their amplitudes.

%\begin{figure}[th!]
% \includegraphics[width=.77\textwidth]{zoom_eps_00145_omega.pdf}
% \caption{zoom of }
% \label{FigET_nuZ}
%\end{figure}

%\begin{figure}[th!]
% \includegraphics[width=.88\textwidth]{power_nu_omega.pdf}
% \caption{Power spectrum of the Jacobi neuronal model from Eq.\eqref{power_sp} for increasing values of $\lambda$, $\epsilon=10^{-4}$ and $\omega_{IN}=0.05$. Peaks in correspondence of certain frequencies are clearly visible.
% For strong excitatory input rates the power spectrum shows an almost deterministic firing behaviour, with a strong peak related to the frequency of $b$.}
% \label{FigPower}
%\end{figure}
%

\begin{figure}[th!]
 \includegraphics[width=.49\textwidth]{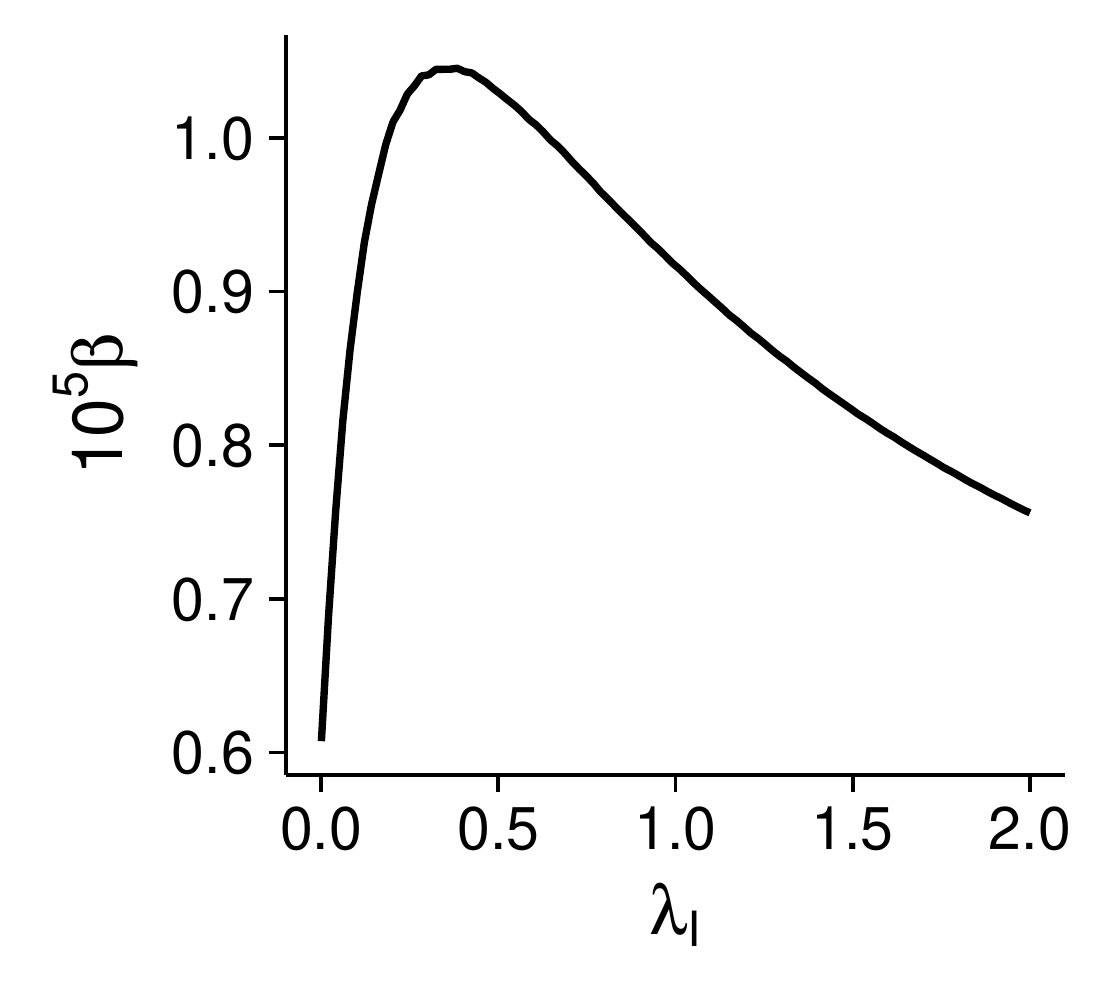}
 \includegraphics[width=.49\textwidth]{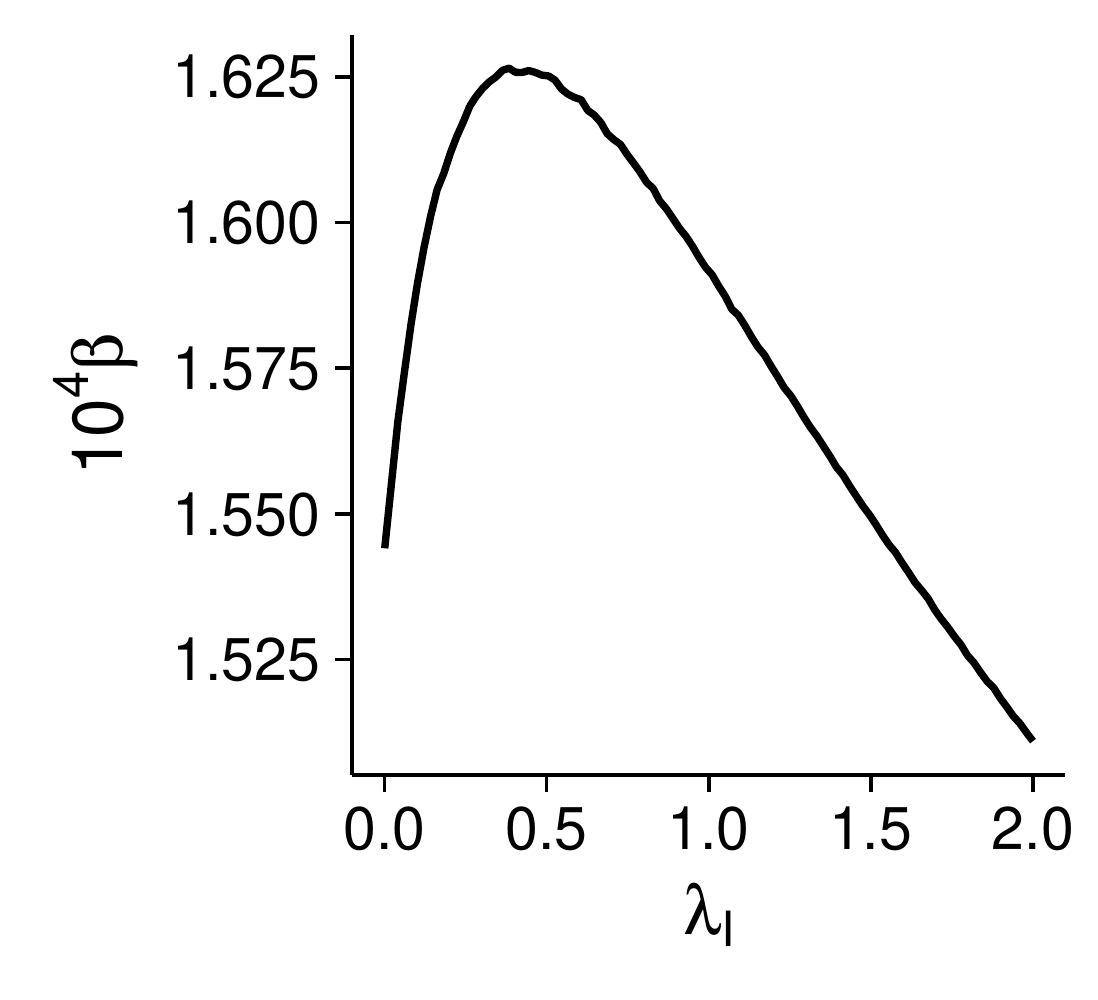}
 \caption{Degree of coherence as a function of $\lambda_I$ for   $\epsilon=0.0145$, $\lambda_E=0.15$ ms$^{-1}$, $\tau=5.8$ ms (left); $\epsilon=0.025$, $\lambda_E=0.34$ ms$^{-1}$, $\tau=3$ ms (right). The other parameters are the same as in Fig.\ref{FigET_nu}.
For $\lambda_I>0.5$ ms$^{-1}$ we observe the classical result: the coherence decreases for increasing inhibition. For $\lambda_I<0.5$ ms$^{-1}$ and for relatively small values of the rate of arrival of excitatory inputs, the inhibition can increase the degree of coherence (even if the increment is small in terms of absolute values). We see that the coherence can be maximized by an optimum value of the inhibitory input rate (in this particular case around $0.5$ ms$^{-1}$).
  The lines are not straight due to small numerical errors in the evaluation of the power spectrum. 
  We choose small $\lambda_E$ because we want weak signals and the input $b$ depends linearly on $\lambda_E$ (see Eq.\eqref{alphabeta}). }
 \label{FigCR}
\end{figure}

The increasing of the excitatory rate increases generally the coherence of the spike train \cite{lind2004}. 
The same happens for the Jacobi model (figure not shown).
The results are generally the opposite for  increasing inhibitory inputs, for which the corresponding neuronal output exhibits a decreased coherence.
The degree of coherence of the Jacobi neuronal model as a function of $\lambda_I$ is shown in Fig.\ref{FigCR}.
Here, for relatively small values of the excitatory rate ($\lambda_E=0.15$ ms$^{-1}$ or $\lambda_E=0.34$ ms$^{-1}$), the inhibition increases the degree of coherence. We observe its maximum 
in the classical bell-shape of the SNR,  typical of the stochastic resonance.

Finally we notice that enhanced coherence is achieved for values of the parameters for which the $D_{eff}$ shows a maximum (see for instance Fig.\ref{FigET_nu} right-panel and Fig.\ref{FigCR} for $\lambda_E=0.15$ ms$^{-1}$). The same effect is observed for $\epsilon=0.025$ (figure not shown).

%%%%%%%%%%%%%%%%%%%%%%%%%%%%%%%%%%%%%%%%%%%
%%%%%%%%%%%%%%%%%%%%%%%%%%%%%%%%%%%%%%%%%%%
\section{Discussion}
%%%%%%%%%%%%%%%%%%%%%%%%%%%%%%%%%%%%%%%%%%%
%%%%%%%%%%%%%%%%%%%%%%%%%%%%%%%%%%%%%%%%%%%
\label{section_5}
At low noise intensities and not sufficient excitation to evoke a spike, the membrane depolarization evolves mainly around the resting potential, with an occasional threshold crossing. Conversely, at large noise regimes, the membrane potential fluctuations are  dominated by the noise, the firings are frequent and highly irregular. Coherence resonance  refers to the phenomenon that occurs at  intermediate intensities of the noise and for which the firings become more regular than at the low and high noise intensity scenarios \citep{gammaitoni,pikovsky,   mcdonn,pakdaman}.
In the case of the Jacobi neuronal model, the amplitude of the noise depends linearly on the input rates \cite{DTL_jacobi,lanska94,longtin_bulsara}, so the approach presented in this paper is in the style of coherence resonance, but conceptually different.
Instead of increasing the noise in presence of a weak signal, we  increase the inhibitory rate,  $\lambda_I$, affecting  simultaneously  the noise and the input $\mu(V_E-X_t)+\nu(X_t-V_I)$, cf. Eqs.\eqref{Lanska_before}-\eqref{input_par2}, or, equivalently,  looking at Eq.\eqref{Lanska}, we are manipulating also the coefficient that plays the role of the time constant. 

In the presence of low excitation, the power spectrum of the Jacobi neuronal model is almost flat and it may suggest that the firing activity is practically Poissonian. However, we observe that an increment of the inhibitory rate can enhance the degree of coherence. Even if the effect is small in absolute value (the quantity $S(\omega) - f$ is much smaller than $f$), as far as we know, it has  never been observed before. 
%An explanation can be the following:
%in the Jacobi model the amplitude of the noise depends linearly on the inhibitory rate, but while we increase the noise, we are simultaneously manipulating the input $\mu(V_E-X_t)+\nu(X_t-V_I)$ (cf. Eq.\eqref{alphabeta}), or equivalently,  looking at Eq.\eqref{Lanska}, we are changing the parameter that plays the role of the time-constant.
Another  property of the Jacobi model discussed here concernes the diffusion coefficient, $D_{eff}$.
It is a commonly accepted fact that the occurrence of a minimum in the diffusion coefficient vs noise intensity is a strong manifestation of coherence resonance \cite{lind2004,lindner2002,guantes}.
In the case of the Jacobi neuronal model, enhanced coherence is observed  for values of the parameters corresponding to a maximum in the diffusion coefficient vs inhibitory rate.
This suggests that, as pointed out for the $\textrm{CV}$ \citep{lind2004,coherence}, also the
$D_{eff}$ should not be used as the only indicator of coherence resonance.

These counterintuitive results are consequences of the presence of the reversal potentials in the model equation and we speculate that the same phenomena can be observed in other models with multiplicative noise, like the Feller or the IGBM \cite{DLP_chaos}.
The extension of these findings to the entire class of models with multiplicative noise will be the subject of a future work.

%%%%%%%%%%%%%%%%%%%%%%%
%%%%%%%%%%%%%%%%%%%%%%%%
\section*{Acknowledgments}
This work was supported by 
the Joint Research Project between Austria and the Czech Republic through Grant No. CZ 19/2019, by the Institute of
Physiology RVO:67985823 and by the Czech Science Foundation
project 17-06943S.

\end{document}